\documentclass[aip,jcp,reprint]{revtex4-1}

\usepackage{amsmath}
\usepackage{amsfonts}
\usepackage{url}
\usepackage{bm}
\usepackage{bbold}
\usepackage{graphics}
\usepackage{paralist}

\newcommand{\bs}{\boldsymbol}
\newcommand{\Lowdin}{L\"{o}wdin }

\newcommand{\bra}{\langle}
\newcommand{\ket}{\rangle}

\def \Nw {{N_{\textrm{w}}}}
\def \Eh {{E_{\textrm{h}}}}

\begin{document}

\title{Non-linear biases, stochastically-sampled effective Hamiltonians and spectral functions in quantum Monte Carlo methods}

\author{Nick S. Blunt}
\email{nicksblunt@gmail.com}
\affiliation{University Chemical Laboratory, Lensfield Road, Cambridge, CB2 1EW, U.K.}
\author{Ali Alavi}
\affiliation{University Chemical Laboratory, Lensfield Road, Cambridge, CB2 1EW, U.K.}
\affiliation{Max Planck Institute for Solid State Research, Heisenbergstra{\ss}e 1, 70569 Stuttgart, Germany}
\author{George H. Booth}
\affiliation{Department of Physics, King's College London, Strand, London WC2R 2LS, U.K.}

\begin{abstract}
In this article we study examples of systematic biases that can occur in quantum Monte Carlo methods due to the accumulation of non-linear expectation values, and approaches by which these errors can be corrected. We begin with a study of the Krylov-projected FCIQMC (KP-FCIQMC) approach, which was recently introduced to allow efficient, stochastic calculation of dynamical properties. This requires the solution of a sampled effective Hamiltonian, resulting in a non-linear operation on these stochastic variables. We investigate the probability distribution of this eigenvalue problem to study both stochastic errors and systematic biases in the approach, and demonstrate that such errors can be significantly corrected by moving to a more appropriate basis. This is lastly expanded to include consideration of the correlation function QMC approach of Ceperley and Bernu, showing how such an approach can be taken in the FCIQMC framework.
\end{abstract}

\maketitle

\section{Introduction}
\label{sec:intro}

The introduction of the full configuration interaction quantum Monte Carlo (FCIQMC) method\cite{Booth2009,Cleland2010,Cleland2011,Spencer2012} has subsequently seen a large number of new quantum Monte Carlo (QMC) methods introduced by various groups, making use of an FCIQMC-like spawning procedure. Such approaches include, coupled cluster Monte Carlo (CCMC)\cite{Thom2012,Scott2017,Neufeld2017}, density matrix quantum Monte Carlo\cite{Blunt2014,Malone2016}, model space quantum Monte Carlo (MSQMC)\cite{Ten-no2013, Ohtsuka2015, Ten-no2017}, and recently driven-dissipative quantum Monte Carlo (DDQMC)\cite{Nagy2018}, to name only a few.

In some cases, these QMC methods are stochastic adaptations of previously-existing deterministic algorithms. Such stochastic adaptations can offer several advantages, perhaps most importantly that stochastic sampling often allows for reduced storage requirements compared to deterministic equivalents, allowing accurate study of extremely large systems in many cases.
However, some significant care is required in reformulating a deterministic algorithm as a stochastic one. Many deterministic methods make use of complicated non-linear operations, often including poorly-conditioned problems. In such cases, it would be potentially careless to assume that a given deterministic method can be converted to a Monte Carlo method in a straightforward manner.

Specifically, when estimating a desired quantity, $f(x)$ (for some other underlying quantity $x$), QMC methods require a large amount of averaging to reduce stochastic errors. For a linear function, $f(x)$ can be estimated by averaging $f(\hat{q})$, where $\hat{q}$ is a random variable being sampled by the QMC method (with $E[\hat{q}] = x$). However, for a non-linear function, it is well know that $E[f(\hat{q})] \ne f(E[\hat{q}])$, and so averaging must be performed \emph{before} the function is evaluated, rather than after. In general this will negate the benefits of the sparse Monte Carlo sampling by requiring storage of large parts of the phase space. In FCIQMC, for example, the variable $\hat{q}$ is typically the sampled wave function from a single iteration, and so estimating $E[\hat{q}]$ requires averaging the entire wave function, eventually requiring as much memory as a fully deterministic approach. One must average until $f(E[\hat{q}]) \approx f(x)$ with sufficient accuracy. For some cases this is simple and the QMC approach is successful. For others, it is challenging, or infeasible entirely. A particular issue occurs when the function is ill-conditioned, meaning that any small error in the stochastic estimate will lead to a substantial error in the final result. In such cases, averaging before the function is applied may not even be practical.

Observables in quantum mechanics can always be written as (at most) quadratic expectation values of the wave function. Given the central importance of these {\em pure} expectation values, projector Monte Carlo methods have long sought to compute these quadratic expectation values in an unbiased fashion, and techniques such as forward-walking and reptation Quantum Monte Carlo have emerged with the aim of removing this non-linear bias\cite{Baroni1999a,Wagner2007,Gaudoin07,Rothstein11,Coccia2012a,Russo2012,Rothstein15,Motta2017}. 
An important example is the calculation of a reduced density matrix (RDM), such as the two-particle RDM $\Gamma_{pq,rs}$,
\begin{equation}
\Gamma_{pq,rs} = \bra \Psi | a_p^{\dagger} a_q^{\dagger} a_s a_r | \Psi \ket,
\end{equation}
which is quadratic in the wave function $| \Psi \ket$. Within the FCIQMC approach, early attempts to sample the two-particle RDM in FCIQMC were hampered by the above bias, only slightly improved by partially averaging the wave function prior to evaluating $\Gamma_{pq,rs}$\cite{Booth2012_2}. This was later corrected by the use of the so-called replica trick\cite{Zhang1993,Hastings2010,Overy2014}, such that the two wave functions in the estimate of $\Gamma_{pq,rs}$ are made statistically independent and the expectation value rewritten as a bilinear functional, formally removing any bias.

While observables are at most quadratic functionals, there often exists the need to compute beyond-quadratic expectation values of stochastically-derived quantities in QMC techniques, such as the computation of entanglement measures, or the sampling and subsequent diagonalization of effective Hamiltonians. This solution to a stochastically-sampled eigenvalue problem is a highly-nonlinear operation, and so there is potential for non-linear biases, and care must be taken. It is these `beyond-quadratic' operations which are the focus of this work, where the above approaches for pure expectation values are not appropriate. 

Direct solution of stochastically-derived eigenvalue problems is not uncommon in QMC methods. Perhaps most notably, the linear method\cite{Umrigar2007, Toulouse2014, Zhao2016, Shea2017}, regularly used to optimize wave function parameters in variational Monte Carlo\cite{McMillan1965, Foulkes2001, Neuscamman2012, Neuscamman2016} (both in continuum and discrete spaces), requires solution of a Hamiltonian eigenvalue problem in a space spanned by the wave function and its first parameter derivatives. In other examples of non-linear operations, FCIQMC has been recently used to perform complete-active space self-consistent field (CASSCF) calculations\cite{Thomas2015_3, Manni2016}, requiring non-linear optimization of orbital coefficients, with no resulting difficulties encountered thus far. The MSQMC method of Ten-no\cite{Ten-no2013, Ohtsuka2015, Ten-no2017} obtains excited states by solving a stochastically-sampled eigenvalue problem, once again with no difficulties reported and highly accurate results. Furthermore, QMC techniques have been used to obtain low-energy effective Hamiltonians for correlated materials\cite{Wagner15}. 

Perhaps the most challenging quantities to sample in QMC (yet among the most highly-sought after) are dynamical properties, specifically many-body Green's functions\cite{AltlandSimons}. FCIQMC, diffusion Monte Carlo (DMC)\cite{Foulkes2001} and other projector QMC methods perform imaginary-time evolution, allowing access to imaginary-time Green's functions as quadratic expectation values which can be sampled with the techniques above. However, the transformation from imaginary-time to real-time (or the frequency domain) is highly non-linear and ill-conditioned\cite{Jarrell1996}, and so challenging to perform in the presence of noise. This transformation is typically performed by maximum-entropy methods\cite{Bryan1990, Silver1990_1, Silver1990_2, Gubernatis1991, Caffarel1992} which, despite sometimes being accurate, are unsatisfactory in general.

As a further example, we recently introduced a QMC approach to estimate many-body Green's functions (and finite-temperature and excited-state properties in general), denoted Krylov-projected FCIQMC (KP-FCIQMC)\cite{Blunt2015_2}. This method can be loosely characterized as a stochastic version of the Lanczos method, and therefore allows access to dynamical and finite-temperature properties in analogy with dynamical and finite-temperature Lanczos. Although Lanczos-type algorithms have been performed with QMC previously\cite{Caffarel1991}, we are unaware of their use to this extent, specifically in the calculation of spectral or finite-temperature properties. KP-FCIQMC was applied to study one and two-particle Green's functions in a one-dimensional Hubbard model, showing that essentially-exact Lanczos results can be reproduced, particularly in low-frequency regions. However, subtle features (particularly at high frequencies) were difficult to reproduce. Clearly, there is an issue with certain eigenvalue problems depending on their nature, with poor conditioning being an obvious potential problem. Given the significant utility of being able to solve such problems, it is worth investigating and discussing such issues, with the KP-FCIQMC approach constituting the exemplar approach for these investigations.

In Section II we re-introduce the KP-FCIQMC method; the theory of many-body Green's functions is introduced in Section III, including a description of their calculation by KP-FCIQMC. In Section IV we apply this approach to one-dimensional Hubbard models in both the weak and intermediate-coupling regimes, including investigation of probability distributions of the solutions; Section V presents a theoretical model to explain these errors, and so demonstrates a solution using trial wave functions. Section VI extends this idea to the correlation function QMC approach of Ceperley and Bernu\cite{Ceperley1988}, which can partly resolve such biases, although systematic errors grow eventually. In Section VII we discuss how excited-state FCIQMC uses orthogonalization to overcome issues of previous sections, and discuss similarities between KP-FCIQMC and dynamical DMRG in the errors observed.

\section{Krylov-projected FCIQMC}

\subsection{Defining the Krylov subspace}

The Krylov-projected FCIQMC (KP-FCIQMC) method is essentially a stochastic adaptation of the Lanczos method (and other Krylov subspace methods in general). In the Lanczos method, one builds the Hamiltonian eigenvalue problem in the subspace spanned by
\begin{equation}
K = \{ |\psi_0\ket, \; \hat{H}|\psi_0\ket, \; \hat{H}^2|\psi_0\ket, \; \ldots, \; \hat{H}^{N_K-1}|\psi_0\ket \},
\label{eq:lanczos_subspace}
\end{equation}
where $|\psi_0\ket$ is some initial state (that can be varied depending on the quantity desired) and $\hat{H}$ is the Hamiltonian operator. The states spanning $K$ are ``Lanczos vectors'', or for a more general subspace, ``Krylov vectors''. We use the latter terminology, and refer to the span of $K$ as the Krylov subspace. Typically, one constructs the Hamiltonian eigenvalue problem in this subspace (which, for this particular subspace, can be put in an efficient tridiagonal form). In deterministic approaches, the Krylov vectors are orthonormalized (implicit in the tridiagonal form), though this does not alter their span, but improves the efficiency and numerical stability of the algorithm.

In the FCIQMC method, we sample the vectors $(\mathbb{1} - \Delta\tau \hat{H})^n |\psi_0\ket$, where $n$ labels the iteration. Thus, the Krylov subspace we work with in KP-FCIQMC is
\begin{multline}
K = \{ |\psi_0\ket, \; (\mathbb{1} - \Delta\tau \hat{H})^{n_1} |\psi_0\ket, \; (\mathbb{1} - \Delta\tau \hat{H})^{n_2} |\psi_0\ket, \\
       \; \ldots, \; (\mathbb{1} - \Delta\tau \hat{H})^{n_{(N_K-1)}}|\psi_0\ket \},
\end{multline}
where $n_l$ labels the FCIQMC iteration at which the $l$'th Krylov-vector is sampled. If both $\Delta\tau$ and $n_{l+1} - n_l$ are small then the vectors will be similar, potentially leading to near-linear dependencies and poor conditioning. In practice, we therefore chose $n_l$ such that Krylov vectors are chosen more frequently at first (as the wave function varies rapidly, and high-energy states are sampled), and less frequently as the ground state is approached.

For notational convenience, we label the $l$'th Krylov vector $| \psi_l \ket \equiv (\mathbb{1} - \Delta\tau \hat{H})^{n_l}|\psi_0\ket$, with $|\psi_0\ket$ the initial vector and $| \psi_{N_K-1} \ket$ the final vector, with $N_K$ Krylov vectors in total.

\subsection{Estimating the subspace Hamiltonian and overlap matrices}

After deciding at which iterations of the FCIQMC algorithm Krylov vectors are to be sampled, the task is to perform those iterations and sample the Hamiltonian and overlap matrix between these vectors. It should be remembered that the `vectors' in this approach are in practice given by sparse, stochastic walker distributions over instantaneously occupied determinants.
The overlap matrix is simple, as we store all Krylov vectors next to each other in an array (containing only determinants sampled in at least one Krylov vector, reducing memory requirements significantly). Thus, calculating all overlap elements requires a dot product between each pair of vectors in this array (naturally parallelized as the array storage is already distributed). Thus the overlap matrix is calculated exactly for each pair of vectors. However, each Krylov vector is stochastically sampled, so noise appears in the overlap matrix regardless.

Estimation of the Krylov-projected Hamiltonian is more challenging, but can be formally sampled using the same spawning dynamics as in standard FCIQMC. To calculate $H^K_{ij} \equiv \bra \psi_i | \hat{H} | \psi_j \ket$ (where $\bs{H}^K$ indicates the Hamiltonian projected into the Krylov subspace), one cycles through each determinant in $ |\psi_i \ket$ and performs FCIQMC spawning to sample determinants in $ | \psi_j \ket$. For some small systems, it is possible to calculate $\bs{H}^K$ exactly (although we again emphasize that Krylov vectors are stochastically sampled, so that stochastic errors remain).

As discussed in the introduction, the calculation of a quantity like $\bra \psi_i | \hat{H} | \psi_j \ket$ is biased if $\bra \psi_i |$ and $| \psi_j \ket$ are sampled from the same QMC simulation, because $E[\hat{q}_i \hat{q}_j] \ne E[\hat{q}_i] E[\hat{q}_j]$, for random variables $\hat{q}_i$ and $\hat{q}_j$, unless they are uncorrelated. This is resolved by using replica sampling, where two FCIQMC simulations are performed simultaneously and independently of each other, removing correlation and therefore bias. We therefore perform two FCIQMC simulations independently, with the first used to sample the `bra' Krylov vectors, $\bra \psi_i |$, and the second used to sample `ket' Krylov vectors, $| \psi_j \ket$, as used in estimation of both $\bra \psi_i | \psi_j \ket$ and $\bra \psi_i | \hat{H} | \psi_j \ket$ elements.

Thus, using replica sampling, estimates of $\bs{H}^K$ and $\bs{S}^K$, are essentially unbiased. Perhaps the only cause of ``bias'' (where ``bias'' here refers to a systematic discrepancy from an otherwise-identical deterministic calculation) is the use of a shift for population control, which typically leads to a negligible error. When studying probability distribution functions later, we keep the shift constant throughout, to remove any theoretical discrepancy in the estimates of $\bs{H}^K$ and $\bs{S}^K$.

\subsection{Comparison to the Lanczos method}

We briefly compare the above approach with the traditional Lanczos method. Firstly, we note that the subspace spanned is only formally the same as that in the Lanczos method, Eq.~(\ref{eq:lanczos_subspace}), when a Krylov vector is sampled at \emph{every} FCIQMC iteration ($n_l = l$). However, this is not necessary. The only requirement is that the space sampled is sufficient to span the ``important'' features of the desired quantity. For example, in calculating spectral properties, it is important that the Krylov vectors contain significant contributions from eigenstates with large amplitudes in the spectrum.

More significantly, one may wonder why the Krylov vectors used are not orthogonalized. In the Lanczos algorithm, the subspace Hamiltonian takes exactly a tridiagonal form, which means that only three Lanczos vectors need storing, and also improves numerical stability. For KP-FCIQMC the situation is somewhat different. Firstly, performing this orthogonalization procedure could introduce a bias into the calculation of $\bs{H}^K$ and $\bs{S}^K$, compared to a deterministic equivalent, because the orthogonalization operation is non-linear. We have recently introduced the excited-state FCIQMC method, where orthogonalization is used and this potential bias is not observed. Nonetheless the situation here is rather different, as subsequent FCIQMC vectors are nearly identical, whereas in excited-state FCIQMC the orthogonalized vectors are approximately orthogonal beforehand, resulting in only a small change from orthogonalization. Secondly this leads into a practical issue, as orthogonalizing a vector with respect to all previous Krylov vector mostly reduces it to zero, which in a QMC algorithm means killing almost all walkers. In traditional Lanczos one need not ever actually perform the orthogonalization, but in a stochastic setting with approximate vectors, this is not the case, and $\bs{H}^K$ would never be exactly tridiagonal regardless.

\subsection{Solution of the subspace eigenvalue problem}

The subsequent subspace eigenvalue problem can be solved by usual methods. This problem can be written as
\begin{equation}
\bs{H}^K \bs{\psi}^K = \epsilon \bs{S}^K \bs{\psi}^K.
\end{equation}
This is solved with a standard canonical \Lowdin orthogonalization procedure, transforming to an orthonormal basis with
\begin{align}
\bs{\psi}^{\textrm{K}} &= \bs{U} \bs{D}^{-1/2} \bs{\psi}^{\textrm{L}}.
\label{eq:lowdin_transformation}
\end{align}
where $\bs{U}$ is the matrix with eigenvectors of $\bs{S}^K$ in its columns, and $\bs{D}$ is the matrix with corresponding overlap-matrix eigenvalues on the diagonal. We call this new basis the \Lowdin basis, using superscript label $L$. The Hamiltonian eigenvalue problem is then
\begin{equation}
\bs{H}^{\textrm{L}} \bs{\psi}^{\textrm{L}} = \epsilon \bs{\psi}^{\textrm{L}},
\end{equation}
with
\begin{equation}
\bs{H}^{\textrm{L}} = \bs{D}^{-1/2} \bs{U}^{\textrm{T}} \bs{H}^{\textrm{K}} \bs{U} \bs{D}^{-1/2}.
\label{eq:lowdin_hamil}
\end{equation}
As discussed above, because many of the Krylov vectors will be nearly linearly dependent, many overlap matrix eigenvalues will be very small, for the exact deterministic problem. In the stochastic problem, the overlap matrix will not take the form of a true overlap matrix, and so will even have negative eigenvalues. We therefore throw away eigenvectors of $\bs{S}^K$ with negative or very small eigenvalues. Typically, we keep $10 - 15$ eigenvectors. These remaining eigenvectors form the final \Lowdin subspace, with dimension denoted $N_L$, such that $N_L < N_K$. Restricting $N_L$ to only $10-15$ means that we only obtain the same number of final eigenvector solutions, which is an approximation to the true spectrum (as it is also for the dynamical Lanczos approach), but is nonetheless useful for many situations, for reasons clarified in the next section.

\section{Many-body Green's functions from KP-FCIQMC}

Green's functions are key observables in many-body condensed matter physics and chemistry, which are directly accessible experimentally, and often allow detailed examination and understanding of important many-body phenomena. As such, their accurate and routine calculation represents a high-priority goal. Unfortunately they are typically difficult to calculate accurately, particularly through QMC methods.

KP-FCIQMC, like dynamical Lanczos, can access two-body Green's functions, but here we focus on the often-studied single-particle (retarded, ground-state) Green's function, defined by
\begin{multline}
G(k,\omega) = \bra \Psi_0 | a_{k\sigma} \: \frac{1}{\omega + \mu - (\hat{H} - E_0) + i\delta} a_{k\sigma}^{\dagger} \: | \Psi_0 \ket \\
 + \bra \Psi_0 | a_{k\sigma}^{\dagger} \: \frac{1}{\omega + \mu + (\hat{H} - E_0) + i\delta} \: a_{k\sigma} | \Psi_0 \ket,
\label{eq:single_particle_corr}
\end{multline}
where $\omega$ is the frequency, $\mu$ is the chemical potential, $a^{\dagger}_{k\sigma}$ is a creation operator for a particle with momentum $k$ and spin label $\sigma$, and $E_0$ and $|\Psi_0 \ket$ are the ground-state energy and eigenvector (in the $N$-particle sector), respectively. $\delta$ is a broadening added to give finite width to poles, but is also related to the rate at which the created quasi-particles decay in an experiment measuring $G(k,\omega)$, so is physically justified. In the non-interacting limit this reduces to the often-quoted expression
\begin{equation}
G(k,\omega) = \frac{1}{\omega + \mu - \xi_k + i\delta},
\end{equation}
for single-particle eigenvalues $\xi_k$. This is the starting point for expanding the interacting Green's function in terms of a self energy, as performed in methods such as dynamical mean field theory (DMFT)\cite{Metzner1989, Georges1996, Kotliar2006}, but we take a more direct approach here. 

The quantity of interest studied here is the spectral function, $A(k,\omega)$, proportional to the imaginary part of the Green's function, typically with $-1/\pi$ as the proportionality factor to ensure normalization,
\begin{equation}
A(\omega) = -\frac{1}{\pi} \Im[G(\omega)].
\end{equation}
Inserting a resolution of the identity in the appropriate $(N+1)$ or $(N-1)$-particle sector gives
\begin{multline}
A(k,\omega) = \sum_i \frac{\delta}{\pi} \frac{ | \bra \Psi_i^{N+1} | \: a_{k\sigma}^{\dagger} \: | \Psi_0 \ket |^2 }{[ \: \omega + \mu - ( E_i^{N+1} - E_0 ) \; ]^2 + \delta^2} \\
 + \sum_i \frac{\delta}{\pi} \frac{ | \bra \Psi_i^{N-1} | \: a_{k\sigma} \: | \Psi_0 \ket |^2 }{[ \: \omega + \mu + ( E_i^{N-1} - E_0 ) \; ]^2 + \delta^2}.
\label{eq:spec_fn}
\end{multline}
where $N+1$ and $N-1$ superscripts denote states in the corresponding sectors in the obvious way. In the limit $\delta \to 0^{+}$, this becomes a sum of delta functions at the energy eigenvalues, with weights given by the transition probabilities $| \bra \Psi_i^{N+1} | \: a_{k\sigma}^{\dagger} \: | \Psi_0 \ket |^2$ and $ | \bra \Psi_i^{N-1} | \: a_{k\sigma} \: | \Psi_0 \ket |^2 $. In the non-interacting limit, this reduces to
\begin{equation}
A(k,\omega) = \delta(\omega + \mu - \xi_k),
\end{equation}
and so $A(k,\omega)$ maps out the single-particle bandstructure.

We also consider the local single-particle density of states (DOS), defined as 
\begin{equation}
A(\omega) = \frac{1}{N} \sum_k A(k,\omega),
\end{equation}
corresponding to a Fourier transform to the real space spectral function at the origin.

We now consider how these quantities can be calculated in KP-FCIQMC. The Lehmann representation expresses the spectral function as a sum over eigenstates, Eq.~(\ref{eq:spec_fn}). The dynamical Lanczos and KP-FCIQMC methods give access to a number of eigenstates equal to the size of the subspace studied, $N_L$ in the case of KP-FCIQMC. This will typically be small compared to the full Hilbert space dimension, and so this may appear unhelpful. However, the subspace is chosen such that the important eigenstates will appear as solutions. The important eigenstates are those with large probability amplitudes, $| \bra \Psi_i^{N+1} | \: a_{k\sigma}^{\dagger} \: | \Psi_0 \ket |^2$ and $ | \bra \Psi_i^{N-1} | \: a_{k\sigma} \: | \Psi_0 \ket |^2 $, as these give the largest contributions to Eq.~(\ref{eq:spec_fn}). Thus, the key is that the Krylov subspace is chosen using $a_{k\sigma}^{\dagger} | \Psi_0 \ket$ as the initial vector to obtain states in the $(N+1)$-particle sector, and $a_{k\sigma} | \Psi_0 \ket$ as the initial Krylov state for the $(N-1)$-particle sector. Thus, those eigenstates with large contributions in Eq.~(\ref{eq:spec_fn}) will have relatively large components in the Krylov subspaces, and may be extracted accurately.

The KP-FCIQMC calculation is started from a perturbed ground state. As such, an FCIQMC calculation is first performed to obatin a stochastic sampling of the ground state. The perturbation is applied from there, and this whole process may be repeated to allow averaging of the (unbiased) $\bs{H}^K$ and $\bs{S}^K$, \emph{before} the (biased) eigenvalue estimate are obtained. This averaging should reduce the \emph{systematic} error in the final eigenvalue estimates and probability amplitudes, thus improving the quality of spectra. This key aspect will be studied in the following sections.

\section{KP-FCIQMC results}
\subsection{One-body Green's functions from KP-FCIQMC}

We study the one-dimensional, periodic $14$-site Hubbard model at half-filling. In our original KP-FCIQMC study, we took $U/t=2$, but here we take $U/t=1$ and $U/t=4$ to study both the nearly-free and intermediate-coupling regimes.

\begin{figure}
\includegraphics{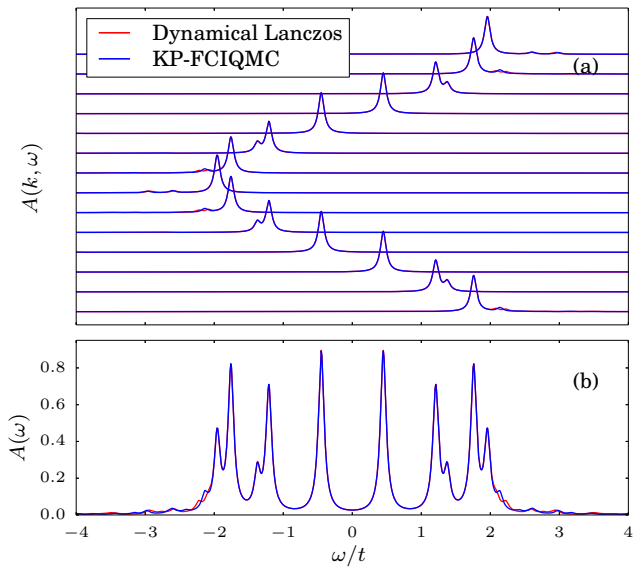}
\caption{(a) $A(k,\omega)$ from $k=-\frac{6}{7}\pi$ (bottom) to $k=\pi$ (top) for the 1D, periodic $14$-site Hubbard model at $U/t=1$, compared to dynamical Lanczos. (b) The local density of states. In this low-coupling regime, the situation is simple and easy to calculate by KP-FCIQMC, with only a small number of low-energy states making significant contributions.}
\label{fig:u1}
\end{figure}

\begin{figure*}
\includegraphics{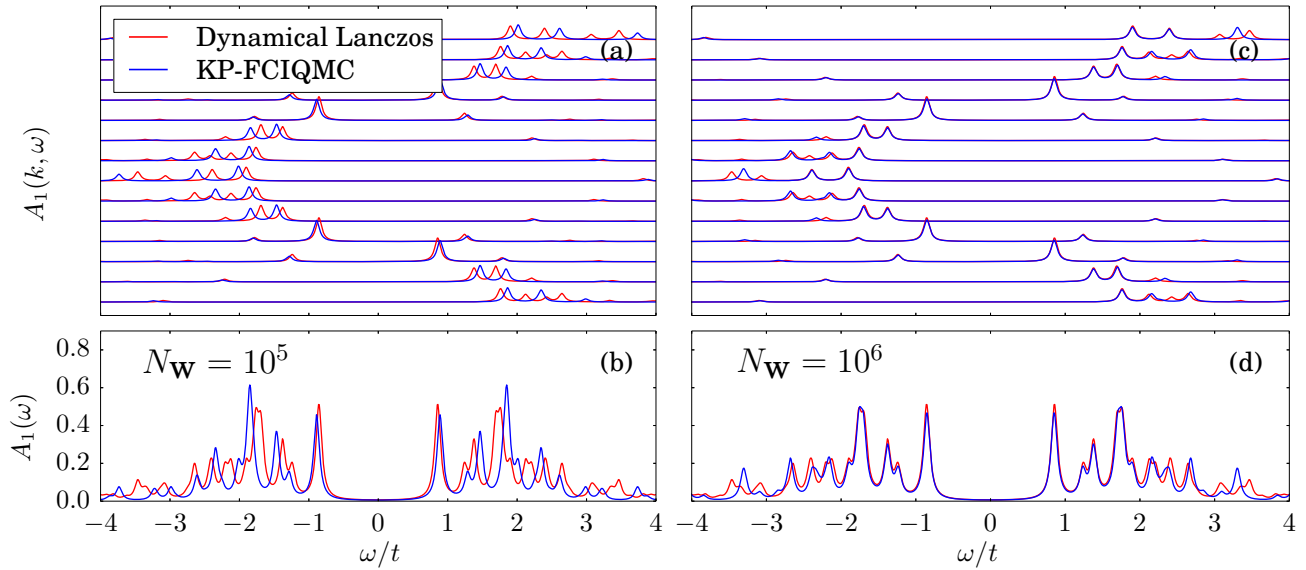}
\caption{Results for the 1D, periodic $14$-site $U/t=4$ Hubbard model, using walker populations, $\Nw$, of both $10^5$, (a) and (b), and $10^6$, (c) and (d). At $\Nw=10^5$, noticeable error exists, which is significantly corrected at $\Nw=10^6$. We believe that this error is largely due to initiator error, which is significant in the Hubbard model at $U/t=4$, although other systematic biases exist. Even at $\Nw=10^6$, errors are more noticeable in the high-frequency regime, which we find to be systematic.}
\label{fig:u4}
\end{figure*}

Results for $U/t=1$ are presented in Fig.~(\ref{fig:u1}). This case is fairly trivial: in this low-coupling regime, the bandstructure is close to the non-interacting bandstructure, where only one or two eigenstates contribute from each $K$ sector. The stochastically-sampled results here accurately reproduce results from dynamical Lanczos. The number of walkers used was $10^5$. The initiator adaptation was used, although initiator error is negligible. The semi-stochastic adaptation was also used with a deterministic space of dimension $|D|=5 \times 10^4$, chosen using the population-based scheme of Ref.~(\onlinecite{Blunt2015}). We set $\delta = 0.05t$. There are only one or two states of significance to sample in each $K$ sector, which are well-sampled in the Krylov vectors, and effectively no bias is noticeable.

In Fig.~(\ref{fig:u4}), the same system is studied but with $U/t=4$. Here the situation is more challenging, with a larger number of eigenstates making a significant contribution, and low-energy states generally making a smaller contribution. We use the same parameters as for the $U/t=1$ case, but study walker populations, $\Nw$, of both $10^5$ and $10^6$. At $\Nw=10^5$ there are clear errors in the whole spectrum. This could be due to a number of factors, but we believe that initiator error is the biggest issue, as initiator error is significant even in ground-state FCIQMC calculations. This is largely resolved at $\Nw=10^6$, where KP-FCIQMC results are in good agreement with dynamical Lanczos.

This shows that the stochastic Krylov-space approach presented here can be successful and accurate, even in the intermediate coupling regime. Because of the sparse sampling (and therefore relatively-low memory requirements), this approach is possible for systems beyond those treatable by dynamical Lanczos (although in this particular example, the density matrix renormalization group (DMRG) method would be a more efficient approach due to its one-dimensional nature). Results are particularly accurate at low $|\omega|$, allowing crucial properties around the band gap to be studied.

Nonetheless, it is informative to study the convergence of this approach to the exact results. At $\Nw=10^6$, results are in good agreement with dynamical Lanczos, but error remains, particularly at high $|\omega|$. We have performed multiple repeated calculations on this system and found that results at high frequency are largely unchanging, suggesting a systematic bias rather than statistical error. In the limit of exact sampling, both $\bs{H}^K$ and $\bs{S}^K$ will be obtained exactly, and so will their eigenvalues, but it is informative to study how this convergence comes about.

\begin{figure*}
\includegraphics{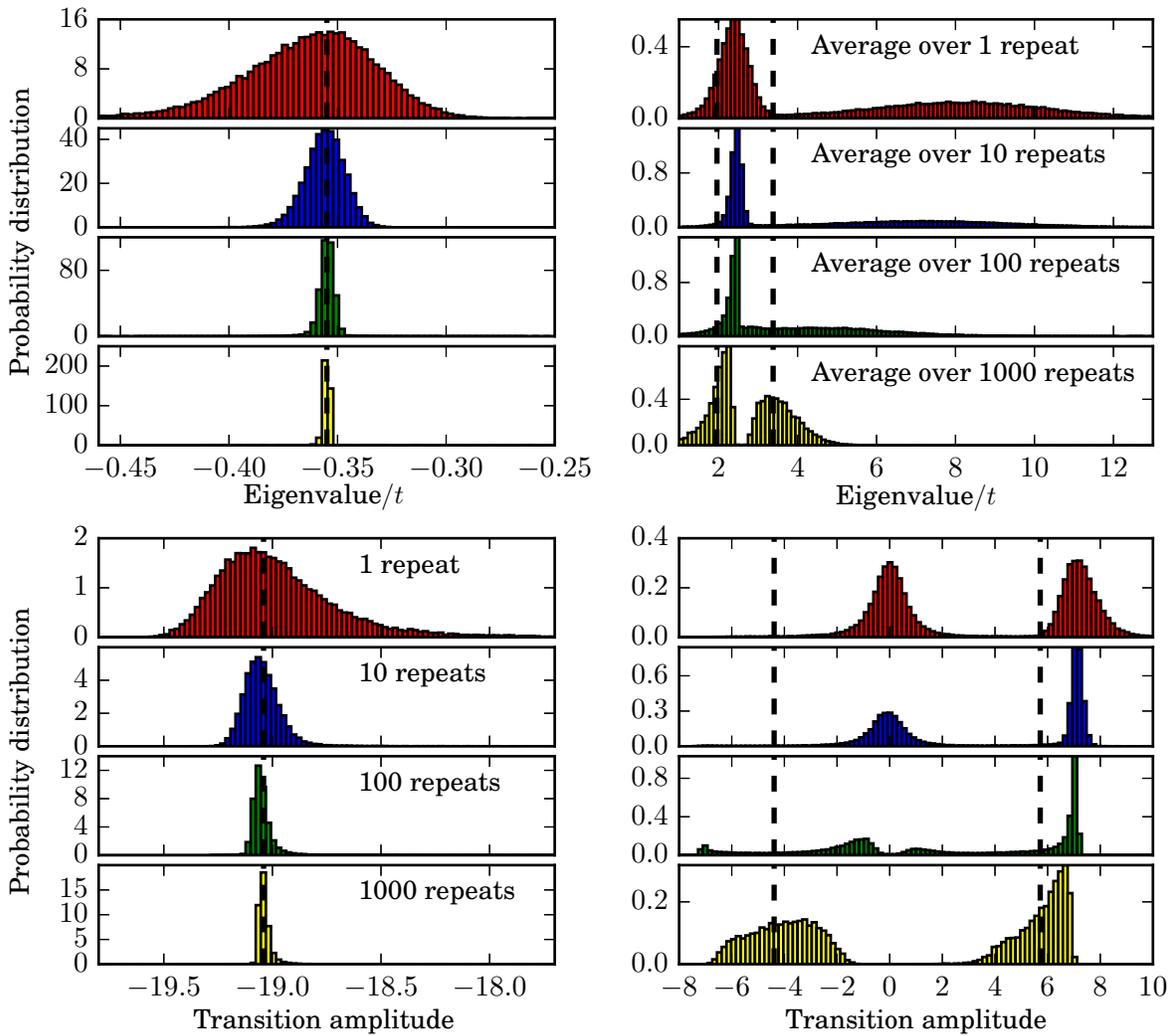}
\caption{Probability distributions for eigenvalues and transition amplitudes from KP-FCIQMC. Dashed lines show exact values. We take a trivial system, the 1D, periodic $6$-site Hubbard model at half-filling and $U/t=4$, in the $K=2\pi/3$ sector. This allows investigation of probability distributions for solutions of stochastically-sampled eigenvalue problems. We aim to study only the $3$ dominant eigenvalues in the spectrum. Results on the left are for the ground state, and on the right for the two excited states. The stated number of repeats is the number of repeats over which $\bs{H}^K$ and $\bs{S}^K$ are averaged $\emph{before}$ the eigenvalue problem is solved. Since the Hamiltonian and overlap matrices themselves are unbiased, this should reduce biases in eigenvalue solutions. This is indeed found to be the case, with reasonable distributions once averaging over $1000$ repeats is performed. However, for less averaging, significant biases occur for the excited-state estimates (right), in both eigenvalues and transition amplitudes.}
\label{fig:pdfs}
\end{figure*}

\subsection{Probability distributions from stochastically-sampled eigenvalue problems}

To study this problem in the most detail, we investigate the probability distributions functions (PDFs) of the underlying eigenvalue estimates, and also PDFs for transition amplitudes, $\bra \psi_i | a_k^{\dagger} | \Psi_0 \ket$. Probability distributions of QMC estimates are rarely considered; it is usually assumed that such distributions are Gaussian due to the central limit theorem, but this is not clear for solutions of poorly-conditioned problems. However, to construct such a distribution requires performing a large number of repeated calculations (here we perform $\sim 10^4$). We therefore consider a much smaller system: the 1D, periodic $6$-site Hubbard model at $U/t=4$. This somewhat mimics the above case, but now in a system where only a small $\Nw$ is required. We also only consider the $K=2\pi/3$ sector, again mimicking a sector with poor results in the above $14$-site case. We use $\Nw \sim 100$ and a deterministic space of the Hartree--Fock determinant and all single and double excitations.

For this system, there are only three eigenstates with significant contributions to the single-particle spectrum. Therefore, to make the probability distributions clear to view and interpret, we take $N_L=3$, i.e., projecting the eigenvalue problem into a space of dimension $3$, such that a stochastically-sampled $3 \times 3$ eigenvalue problem is obtained. We compare the eigenvalue estimates to those from a completely deterministic KP-FCIQMC calculation, but with the same projection into a $3 \times 3$ problem performed, so that exactly the same eigenvalues will be obtained in the limit of infinite averaging in the stochastic case.

We perform $\sim 10^4$ repeated KP-FCIQMC calculations, each with a different random number generator (RNG) seed. Within each of these $10^4$ repeats, we average $\bs{H}^K$ and $\bs{S}^K$ over either $1$, $10$, $100$ or $1000$ repeats, \emph{before} the eigenvalue problem is solved. As more averaging of the unbiased $\bs{H}^K$ and $\bs{S}^K$ is performed, any bias in the eigenvalue estimates should be reduced. The shift is kept constant to avoid any theoretical population-control discrepancies.

The constructed PDFs are presented in Fig.~(\ref{fig:pdfs}). Results on the left are for the ground-state solutions, while those on the right are for the two excited states. Results at the top show PDFs for the energy eigenvalues themselves, while those at the bottom are PDFs for the transition amplitudes, $\bra \psi_i | a_k^{\dagger} | \Psi_0 \ket$.
PDFs for the ground state are sensible for all levels of averaging performed. There is a slight skew occurring when no averaging is performed (i.e., $1$ `repeat'), although even in this case the PDF mean is approximately correct.

In contrast, PDFs for the two excited states show results in significant error. For energy eigenvalues (top) with no averaging, the first excited state is slightly too high, while the second eigenstate is in error on average by more than $4t$, with a variance so large that results are smeared out entirely. Increasing averaging up to $1000$ repeats does eventually bring results to be distributed about approximately the correct values, although even then with a strong skew.

Transition amplitudes for excited states are also in significant error until significant averaging is performed. For low-averaging, one of the excited states has a transition amplitude which is approximately $0$ (while the other is too large, such that two peaks are merged into one), which does not begin to resolve until averaging over $100$ repeats, and not satisfactorily until averaging over $1000$ repeats.
Clearly, even though this is a simple system, significant biases occur in the eigenvalue estimates of the excited states, although ground-state estimates are accurate. While the number of walkers is similarly small with the system size, this should act as a warning for the issues that can occur in stochastically-sampled eigenvalue problems, even in simple cases. Despite this, given the utility of such solutions in QMC methods, this issue is usually not so severe.

\section{Analysis of error for a two-state model}

We now consider a theoretical model in only two dimensions, where the source of this significant bias can be identified.

Suppose we sample two Krylov vectors of the form
\begin{equation}
| K_0 \ket = | \psi_0 \ket \; \; \; \; \; \; \; \; \; \; | K_1 \ket = | \psi_0 \ket + \delta | \psi_1 \ket,
\end{equation}
where $| \psi_0 \ket$ and $| \psi_1 \ket$ are orthonormal solutions which we seek to obtain, and $\delta$ is some small and positive number. This mimics what happens in real simulations where there is one vector (the ground state) making up a relatively large component of all Krylov vectors. Then the eigenvalue problem is
\begin{equation}
\begin{pmatrix}
a & a \\
a & a + \delta^2 b
\end{pmatrix}
\begin{pmatrix}
x_0 \\
x_1
\end{pmatrix}
= \epsilon
\begin{pmatrix}
1 & 1 \\
1 & 1 + \delta^2
\end{pmatrix}
\begin{pmatrix}
x_0 \\
x_1
\end{pmatrix}
,
\end{equation}
where $a = \bra \psi_0 | H | \psi_0 \ket$, $b = \bra \psi_1 | H | \psi_1 \ket$ and $\bra \psi_0 | H | \psi_1 \ket$ has been defined as $0$ for simplicity.

It can be shown that the eigenvalues of the overlap matrix are (using zero-indexing),
\begin{equation}
D_{00} = 2 + \mathcal{O}(\delta^2) \; \; \; \; \; \; \; D_{11} = \frac{\delta^2}{2} + \mathcal{O}(\delta^4)
\end{equation}
and that the corresponding eigenvectors are
\begin{equation}
\begin{pmatrix}
\frac{1}{\sqrt{2}} \\
\frac{1}{\sqrt{2}}
\end{pmatrix}
+ \mathcal{O}(\delta^2) \; \; \; \; \; \; \; \; \; \;
\begin{pmatrix}
\frac{-1}{\sqrt{2}} \\
\frac{1}{\sqrt{2}}
\end{pmatrix}
+ \mathcal{O}(\delta^2).
\end{equation}
The Hamiltonian in the \Lowdin basis (the final orthonormal basis) is
\begin{equation}
\bs{H}^{\textrm{L}} = 
\begin{pmatrix}
a + \mathcal{O}(\delta^2) & \mathcal{O}(\delta) \\
\mathcal{O}(\delta) & b + \mathcal{O}(\delta^2)
\end{pmatrix}
.
\end{equation}
The eigenvalues of the exact $\bs{H}^{\textrm{L}}$ matrix are $a$ and $b$, as expected by construction.

If this eigenvalue problem were sampled by KP-FCIQMC, then the stochastic estimate of the Krylov-space Hamiltonian could be written
\begin{equation}
\bs{H}^{\textrm{K}} + \bs{\eta},
\end{equation}
where $\bs{H}^{\textrm{K}}$ is the \emph{exact} Krylov-space Hamiltonian, and $\bs{\eta}$ is the error matrix (and symmetry is enforced on the Hamiltonian estimate, so that $\eta_{10} = \eta_{01}$). If the transformation matrix from the Krylov to the \Lowdin basis is denoted $\bs{T}$, then the stochastic estimate of the Krylov-projected Hamiltonian transforms as
\begin{equation}
\bs{T}^{\textrm{T}} (\bs{H}^{\textrm{K}} + \bs{\eta}) \bs{T} = \bs{T}^{\textrm{T}} \bs{H}^{\textrm{K}} \bs{T} + \bs{T}^{\textrm{T}} \bs{\eta} \bs{T}.
\end{equation}
The first term is the desired $\bs{H}^{\textrm{L}}$. We would like second term to equal $\bs{0}$. For finite stochastic error, however, this term can be shown to equal (to leading order in $\delta$)
\begin{equation}
\bs{T}^{\textrm{T}} \bs{\eta} \bs{T} =
\begin{pmatrix}
\frac{1}{4}(\eta_{00} + 2\eta_{01} + \eta_{11}) & \frac{1}{2\delta} (\eta_{11} - \eta_{00}) \\
\frac{1}{2\delta} (\eta_{11} - \eta_{00}) & \frac{1}{\delta^2} (\eta_{00} - 2\eta_{01} + \eta_{11})
\end{pmatrix}
.
\end{equation}
Therefore, $H_{11}^{\textrm{L}}$ has a large error when $\delta$ is small. Roughly speaking, to obtain an accurate estimate of the Hamiltonian in the final orthonormal basis, the Hamiltonian in the Krylov basis must have errors $\eta_{ij}$ smaller than $\delta^2 \approx D_{11}$. This makes more rigorous the intuitive notion that eigenvectors with a small component in the Krylov vectors require a similarly small associated stochastic error for accurate estimation, quickly becoming unreasonable. In this case it is difficult to `extract' the $| \psi_1 \ket$ solution accurately from the noise.

\begin{figure}[t!]
\includegraphics{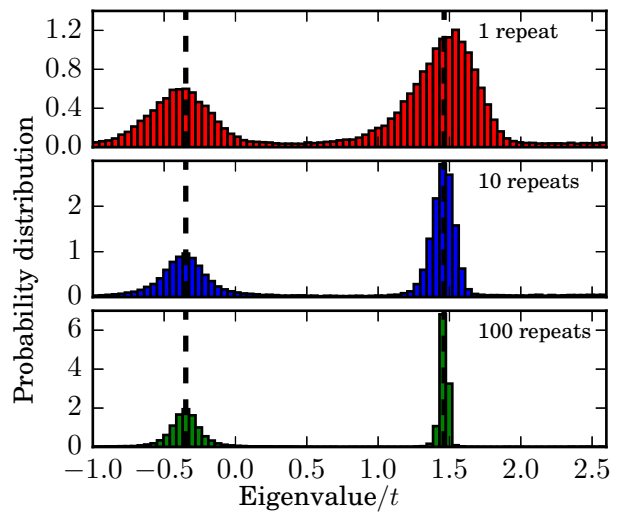}
\caption{Distributions of KP-FCIQMC eigenvalues around the exact ground and first-excited energies (dashed lines) for the same system as Fig.~(\ref{fig:pdfs}), as the number of repeats (over which $\bs{H}^{\textrm{K}}$ and $\bs{S}$ are averaged) is varied. In contrast to Fig.~(\ref{fig:pdfs}), the initial Krylov vector was here taken as the CISD estimate of the first excited state. Both ground and first excited states are obtained with relatively small biases for $1$ repeat, and with no visible bias when averaging over $10$ or more repeats.}
\label{fig:eigv_dist_targ}
\end{figure}

The above analysis suggests that the main source of bias is due to the desired states having small components in the Krylov vectors. It is therefore informative to study different subspaces, where the Krylov vectors are constructed to be similar to the desired excited states.

Such an example is demonstrated in Fig.~(\ref{fig:eigv_dist_targ}). Here we take the same system as that studied in Fig.~(\ref{fig:pdfs}). However, while Fig.~(\ref{fig:pdfs}) took the initial Krylov vector as $a_k^{\dagger} | \Psi_0 \ket$, where $| \Psi_0 \ket$ is the exact ground-state wave function in the $N$-particle sector ($N=6$) (appropriate for constructing the single-particle Green's function), here we take the initial Krylov vector as the CISD estimate to the first excited state in the same final sector ($N=7$). As can be seen, by choosing the subspace more appropriately, the bias is effectively removed in the first excited state estimates. Note that the numerical value for the first excited state is different between Figs.~(\ref{fig:pdfs}) and (\ref{fig:eigv_dist_targ}) - this is because we project into different final $3$-dimensional subspaces in each case. However, this is unimportant; here we are simply assessing biases in the different approaches.

\section{A correlation function QMC approach in FCIQMC}

The approach considered in Fig.~(\ref{fig:eigv_dist_targ}) overcame systematic errors by constructing the subspace in terms of trial wave functions. This approach is also taken in the correlation function quantum Monte Carlo (CFQMC) method of Ceperley and Bernu\cite{Ceperley1988, Bernu1990, Kwon1996, Brown1995, Shumway2001, Meyer2013}, which we now briefly consider within an FCIQMC context. Note that this approach is called correlation function QMC because the sampled $H^{\textrm{C}}_{ij}$ and $S^{\textrm{C}}_{ij}$ (see below) can be viewed as imaginary-time correlation functions. However, in this method one is only interested in eigenvalues and properties of a small number of low-lying excited states, \emph{not} Green's functions, where the entire spectrum is (in principle) obtained. Nonetheless the approach is interesting to consider in relation to the current question of biases and stochastically-sampled eigenvalue problems.

Suppose one has $m$ trial functions, $\left\{ |f_0\ket, |f_1\ket, \ldots, |f_{m-1}\ket \right\}$, for the $m$ lowest energy eigenstates of a system. Improved solutions can be formed by taking linear combinations of the original set of states. The best linear combinations (in a variational sense) are formed by solving the Hamiltonian eigenvalue problem, projected into the space spanned by $\left\{ |f_i\ket \right\}$,
\begin{equation}
\bs{H}^{\textrm{C}} \bs{\phi}_i = \Lambda_i \bs{S}^{\textrm{C}} \bs{\phi}_i,
\end{equation}
where $H^{\textrm{C}}_{ij} = \bra f_i|\hat{H}|f_j \ket$ and $S^{\textrm{C}}_{ij} = \bra f_i|f_j \ket$. The eigenvalue estimates, $\Lambda_i$ are variational by MacDonald's theorem\cite{MacDonald1933} and so will be improved by using more or better-quality trial states, $|f_i\ket$.

Define projected trial states by
\begin{equation}
|f_i(\tau) \ket = \textrm{e}^{-\tau \hat{H}} |f_i \ket.
\end{equation}
If the Hamiltonian eigenvalue problem is evolved with $\tau$, so that $H^{\textrm{C}}_{ij}(\tau) = \bra f_i(\tau)|\hat{H}|f_j(\tau) \ket$, $S^{\textrm{C}}_{ij}(\tau) = \bra f_i(\tau) | f_j(\tau) \ket$ and
\begin{equation}
\bs{H}^{\textrm{C}}(\tau) \bs{\phi}_i(\tau) = \Lambda_i(\tau) \bs{S}^{\textrm{C}}(\tau) \bs{\phi}_i(\tau),
\end{equation}
then since the slowest decaying contributions to each trial state will span the lowest-lying excited states, in principle, the exact energy eigenstates and eigenvalues are retrieved in the limit of large imaginary time, as
\begin{equation}
\lim_{\tau \to \infty} \Lambda_i(\tau) = E_i.
\end{equation}
The above approach is that taken in the CFQMC method.

\begin{figure}[t!]
\includegraphics{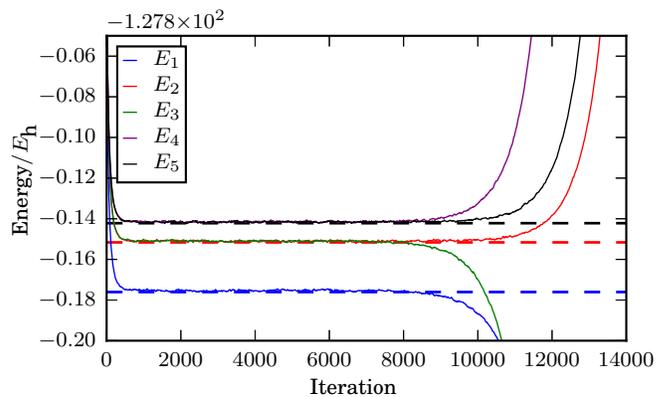}
\caption{FCIQMC energies (from a variational estimator) during propagation for the neon atom in an aug-cc-pVDZ basis, with $2$ core electrons frozen. Simulations begin from the CISD estimates of the five lowest excited states (labelled $E_1-E_5$). Energies converge quite stably to exact energies (dashed lines) for considerable imaginary time, before converging towards the ground state. Note that three states appear to diverge to higher energies, which occurs when the ground state component gains an opposite sign in the two replica simulations, resulting in the denominator of the energy estimator passing through zero. However, the ground state is converged upon eventually (not shown). Note also that there are double degeneracies in two states because the $A_g$ irrep of $D_{2h}$ is used.}
\label{fig:cfqmc_ne}
\end{figure}

\begin{figure}[t!]
\includegraphics{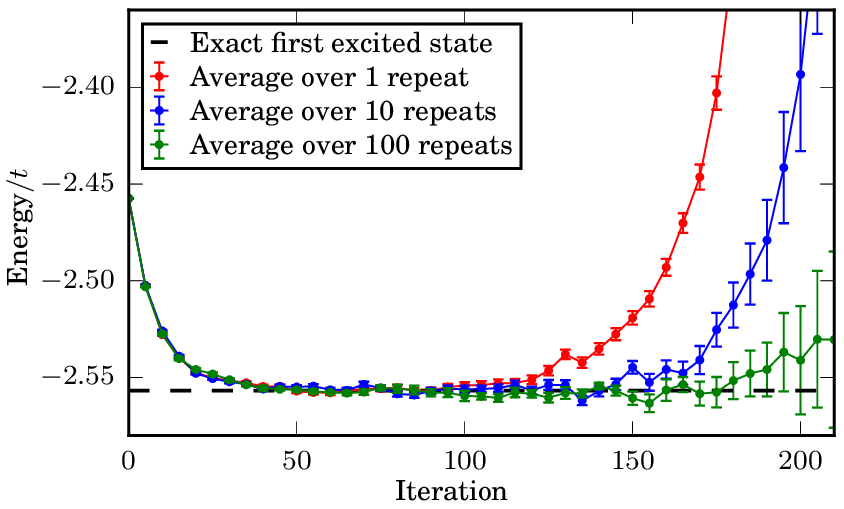}
\includegraphics{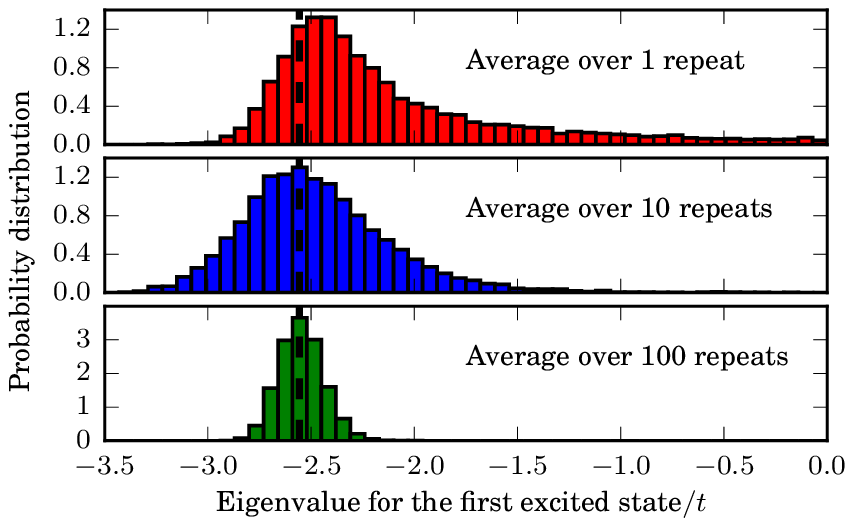}
\caption{Results from the CFQMC procedure for the first excited state, for the one-dimensional, periodic $6$-site Hubbard model at half-filling and $U/t=2$. Simulations began from CISD estimates to the ground and first excited states. $\bs{H}^{\textrm{C}}$ and $\bs{S}^{\textrm{C}}$ were averaged over $1$, $10$ and $100$ repeated simulations \emph{before} eigenvalues were obtained. Top: Mean eigenvalues estimates as a function of imaginary-time, with standard deviations presented as error bars. Eigenvalue bias increases exponentially for large $\tau$. Bottom: PDFs of eigenvalues for the first excited state at iteration $200$ ($\tau=2$).}
\label{fig:cfqmc_eigv_plots}
\end{figure}

However, the approach above is only stable in the complete absence of numerical errors, either due to finite precision, or stochastic noise. This is because for large $\tau$, all states $|f_i(\tau)\ket$ will converge to the ground state, and the trial states will exponentially become linearly dependent, and eigenvalues of the overlap matrix will become exponentially small. Stochastic or numerical errors in $\bs{H}^{\textrm{C}}$ will therefore be greatly magnified, and large biases will result, as for KP-FCIQMC, resulting in a transient estimate of the excited eigenstates in practice. However, it is usually possible to converge far enough to obtain good eigenvalue estimates before such issues become unmanageable.

In the original approach of Ceperley and Bernu\cite{Ceperley1988}, this procedure was performed in real space. However, it is just as simple to perform in an FCIQMC-framework, as there one also performs imaginary-time propagation of initial states. Indeed, the approach is perhaps more convenient here, as it is simple to construct particularly-accurate trial excited states in finite-dimension Hilbert spaces, using standard quantum chemistry methods.

Before applying the full CFQMC approach, we first consider what happens when FCIQMC simulations are performed starting from trial solutions to excited states. This has already been performed in the excited-state FCIQMC approach, but in that approach orthogonalization prevents collapse to the ground state. Here we consider propagation without orthogonalization. An example of this is considered in Fig.~(\ref{fig:cfqmc_ne}), where the Ne atom is considered in an aug-cc-pVDZ basis set for the five lowest excited states, initialized from configuration interaction singles and doubles (CISD) solutions to the states. We use the following energy estimator,
\begin{equation}
E_i(\tau) = \frac{\bra f_i(\tau)| \hat{H} |f_i(\tau) \ket}{\bra f_i(\tau) | f_i(\tau) \ket},
\label{eq:var_est}
\end{equation}
calculated in the same way as in KP-FCIQMC (using the spawning procedure to sample pairs of determinants, and the replica trick). Results of Fig.~(\ref{fig:cfqmc_ne}) show that, perhaps surprisingly, the energy estimates converge accurately towards the exact eigenvalues for over $6000$ iterations before eventually converging to the ground-state energy. One might have expected the collapse towards $E_0$ to happen much quicker, especially starting from fairly basic CISD estimates, even for this simple system. Note that when collapse to the ground state does occur, some energy estimates diverge to higher values, rather than directly to the ground state. This is because the component of the ground state in the two replicas sampling $|f_i(\tau) \ket$ obtain opposite signs in early iterations due to stochastic noise, causing the denominator, $\bra f_i(\tau) | f_i(\tau) \ket$, to pass through $0$ during convergence.
 
The CFQMC procedure then projects $\hat{H}$ into the subspace spanned by these states, and solves the resulting eigenvalue problem to prevent this collapse. An example of this is shown in Fig.~(\ref{fig:cfqmc_eigv_plots}). Here, to allow study of PDFs, we again study the same $6$-site Hubbard model as in Figs.~(\ref{fig:pdfs}) and (\ref{fig:eigv_dist_targ}), but here taking $U/t=2$. We consider the first excited state. While the collapse of the expectation values is prevented in the zero-stochastic-error (and infinite numerical precision) limit, in a stochastic setting the bias on the eigenvalues grows exponentially, as proven by Ceperley and Bernu in their original presentation\cite{Ceperley1988}. This exponential growth of bias is clear here. When performing no averaging, significant error occurs by iteration $200$ ($\tau=2$), although this can be corrected by averaging to reduce noise on $\bs{H}^{\textrm{C}}$ and $\bs{S}^{\textrm{C}}$.

Despite the eventual breakdown, the first excited state is sampled with a near-exact energy for a considerable period of imaginary time before this occurs. This situation is therefore significantly improved compared to that in KP-FCIQMC, due to the use of trial solutions which are already close to spanning the target subspace, resulting in a well-conditioned problem where the first excited state has significant amplitude in the sampled subspace.

\section{Discussion}

\subsection{Excited-state FCIQMC}

The exponentially-growing error in CFQMC with $\tau$ is caused by all states converging to the ground state, resulting in a poorly-conditioned eigenvalue problem. The obvious solution to this is to orthogonalize each state against all lower-energy states, to prevent collapse. This approach was taken in our excited-state extension to FCIQMC\cite{Blunt2015_3}. The projection here takes the form 
\begin{equation}
|f_i(\tau+\Delta\tau)\ket = \hat{O}_i(\tau+\Delta\tau) [\mathbb{1} - \Delta\tau (\hat{H} - S \mathbb{1})] |f_i(\tau)\ket,
\end{equation}
with
\begin{equation}
\hat{O}_i(\tau) = \mathbb{1} - \sum_{j < i} \frac{|f_j(\tau)\ket \bra f_j(\tau)|}{\bra f_j(\tau) | f_j(\tau) \ket}.
\label{eq:orthog_proj}
\end{equation}
Here, the initial ground-state trial wave function, $|f_0\ket$, is evolved exactly as in standard FCIQMC. All higher-energy states follow a similar evolution, but with $\hat{O}_i(\tau)$ applied after each iteration.

With biases in mind, a possible concern in this approach is that the orthogonalization operator $\hat{O}_i(\tau)$ is non-linear in $|f_j(\tau)\ket$, so that biases may occur. This was considered in some detail in our initial presentation. Although we believe that some extreme limits must exist where the non-linear nature leads to undesired results, in practice we have never found this issue to occur and indeed have obtained extremely accurate results in all cases tested. For a test Hubbard model example, any bias was shown to be less than $\sim 10^{-4}-10^{-6}t$ for all states studied. For the stretching of C$_2$ in a cc-pVQZ basis we found agreement with DMRG results to $\sim 10^{-4}\Eh$, with remaining error almost certainly from the initiator approach.

\begin{figure}[t!]
\includegraphics{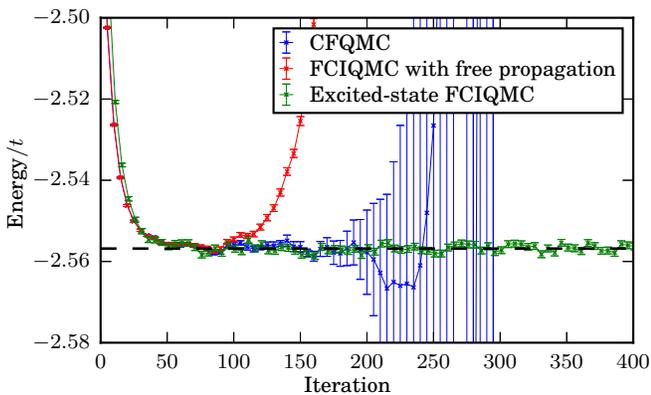}
\caption{Results for the first excited state of the one-dimensional, periodic $6$-site Hubbard model at half-filling and $U/t=2$. Simulations were initialized from CISD wave functions. Free propagation is shown in red, where large errors occur quickly. The CFQMC procedure is shown in blue (here $\bs{H}^{\textrm{C}}$ and $\bs{S}^{\textrm{C}}$ were averaged over $10^3$ repeated simulations before the plotted eigenvalues were obtained), demonstrating improved stability. The excited-state FCIQMC approach is shown in green, where orthogonalization ensures stable convergence to the exact result. Note that under free propagation the ground state is eventually converged upon, although divergence to higher energies is observed here; this occurs for the same reason as in Fig.~(\ref{fig:cfqmc_ne}).}
\label{fig:comparison}
\end{figure}

As a demonstration, we again consider the test case of the one-dimensional $6$-site Hubbard model at half-filling and $U/t=2$, as in Fig.~(\ref{fig:cfqmc_eigv_plots}). Simulations are initialized from CISD wave functions, with the aim of sampling the exact first excited state. A comparison is performed between three cases: free FCIQMC propagation; the CFQMC procedure; and the excited-state FCIQMC procedure, performing orthogonalization against the ground-state FCIQMC wave function. For excited-state FCIQMC results, an RDM-based energy estimator is used (with replica sampling)\cite{Overy2014}, for a relevant comparison. CFQMC is more stable than free propagation, allowing the first excited state to be accurately sampled for a longer period of imaginary time. The orthogonalization-based approach is fully stable. Continuing this simulation for $10^6$ iterations (using $\sim 80$ walkers on average) gives an energy estimate of $-2.55677(23)t$, compared to the exact value of $-2.55683t$, showing no bias to good accuracy.

A similar orthogonalization approach would not be appropriate in real-space projector QMC methods, as the overlap between two stochastically sampled wave functions in real space will be zero. Clearly there are advantages and disadvantages to both real-space and finite-space approaches. The ability to calculate overlaps between two statistically-sampled wave functions, as used in both the excited-state approach and the unbiased sampling of RDMs, has been a large benefit in FCIQMC.
As such, the problem of sampling a small number of low-lying states by FCIQMC has, we believe, been effectively solved by the above orthogonalization approach. However, the task of computing dynamical properties remains a more significant and open challenge.

\subsection{Comparison of KP-FCIQMC with DMRG approaches}

The KP-FCIQMC approach can be compared to similar approaches for calculating dynamical correlation functions in the DMRG framework. Several methods have been attempted by a variety of approaches\cite{Hallberg1995, Ramasesha1998, Kuhner1999, Holzner2011, Ganahl2014}. An approach based on the Lanczos algorithm was attempted by Hallberg\cite{Hallberg1995}, which is directly comparable to the KP-FCIQMC approach. A similar approach has been used recently by Dargel \emph{et al}.\cite{Dargel2011, Dargel2012} This approach was also investigated by K\"{u}hner and White in 1999\cite{Kuhner1999}, who interestingly came to similar conclusions in DMRG as we have in FCIQMC - they state that {\em ``the Lanczos vector method works very well if only the low-energy part of the correlation function is of interest, or if the bulk of the weight is in one single peak.''} This makes evident that these discrepancies are not simply the result of stochastic-type errors; small systematic errors due to other approximations are equally problematic.

\section{Conclusion}

We have investigated examples of sampling non-linear functions in QMC methods, including ill-conditioned problems. KP-FCIQMC allows sampling of dynamical and finite-temperature properties by an approach comparable to the Lanczos method, but with memory limitations removed due to stochastic sampling. For small Hubbard model examples we obtained accurate spectra even in the intermediate-coupling regime, demonstrating the potential of this stochastic Krylov-projected approach, which may be appropriate as an impurity solver for DMFT.

However, in results similar to those obtained from a DMRG-based approach to the spectral Lanczos method\cite{Kuhner1999}, we find that high-energy spectral features are challenging to obtain accurately, particularly for poles with small transition amplitudes. This was demonstrated in a small $6$-site Hubbard model, where probability distributions were obtained for the KP-FCIQMC eigenvalues and transition amplitudes, showing large errors in relatively high-energy states, eventually corrected by performing further averaging. A simple two-state theoretical model was considered to explain this issue, showing that the problem can be traced back to states with small components in the Krylov vectors, leading to an ill-conditioned problem, for which QMC is not best-suited.

By moving to a more appropriate subspace, where a desired excited state has large components in the Krylov vectors, this bias can be largely removed. An example of this is the CFQMC method, which largely resolves such issues for small $\tau$, although biases eventually become significant at large imaginary time.

A large number of stochastic quantum chemistry methods have been formulated in recent years in a similar vein to FCIQMC; we hope that the considerations presented here will be informative, and beneficial in the appropriate formulation of such approaches in the future.

\section{Acknowledgments}

N.S.B. gratefully acknowledges St John's College, Cambridge for funding through a Research Fellowship. A.A. thanks EPSRC for funding under Grant No. EP/J003867/1. G.H.B. gratefully acknowledges funding from the Air Force Office of Scientific Research via grant number FA9550-16-1-0256, and the Royal Society via a University Research Fellowship. The project has received funding from the European Union's Horizon 2020 research and innovation programme under grant agreement No. 759063. We are also grateful to the UK Materials and Molecular Modelling Hub for computational resources, which is partially funded by EPSRC (EP/P020194/1).

\end{document}